\documentclass[runningheads]{llncs}

\usepackage[T1]{fontenc}

\usepackage{graphicx}
\usepackage{booktabs}
\usepackage{tikz}

\usepackage{xcolor}
\usepackage{ulem}
\usepackage{xspace}

\newcommand{\system}{Script\&Shift\xspace} 
\newcommand{\systems}{Script\&Shift's\xspace}

\begin{document}
\title{AI in the Writing Process: How Purposeful AI Support Fosters Student Writing}
\titlerunning{AI in the Writing Process}

\author{Momin N. Siddiqui\inst{1}\orcidID{0000-0003-1874-7789} \and
Vryan Feliciano\inst{2}\orcidID{0009-0005-8802-641X} \and
Roy Pea \inst{2}\orcidID{0000-0001-6301-3536} \and
Hari Subramonyam \inst{2}\orcidID{0000-0002-3450-0447}} 

\authorrunning{M. Siddiqui et al.}

\institute{Georgia Institute of Technology, Atlanta, GA 30332, USA \and
Stanford University, Stanford, CA 94305, USA\\
\email{msiddiqui66@gatech.edu, \{vgfelica, roypea, harihars\}@stanford.edu}}

\maketitle              
\begin{abstract}
The ubiquity of technologies like ChatGPT has raised concerns about their impact on student writing, particularly regarding reduced learner agency and superficial engagement with content. While standalone chat-based LLMs often produce suboptimal writing outcomes, evidence suggests that purposefully designed AI writing support tools can enhance the writing process. This paper investigates how different AI support approaches affect writers' sense of agency and depth of knowledge transformation. Through a randomized control trial with 90 undergraduate students, we compare three conditions: (1) a chat-based LLM writing assistant, (2) an integrated AI writing tool to support diverse subprocesses, and (3) a standard writing interface (control). Our findings demonstrate that, among AI-supported conditions, students using the integrated AI writing tool exhibited greater agency over their writing process and engaged in deeper knowledge transformation overall. These results suggest that thoughtfully designed AI writing support targeting specific aspects of the writing process can help students maintain ownership of their work while facilitating improved engagement with content.

\end{abstract}

\section{Introduction}

Writing --- is not merely an act of text generation but a cognitive process of meaning-making, requiring writers to synthesize and organize ideas, refine arguments, adjust language and framing for different readers, etc. Skilled writers engage in \textit{knowledge transformation}, continuously negotiating between the content (i.e., what to say) and rhetoric (how to say it)~\cite{bereiter2013psychology}. As generative AI is becoming more embedded in writing 
tools~\cite{kim2022learner,lee2024design,siddiqui2025scriptshift,singh2023hide}, emerging research also indicates that these systems may inadvertently \textit{undermine} essential human cognitive processes~\cite{bastani2024generative,singh2025protecting,tarchi2024use}. While large language model (LLM) based writing tools can generate text effortlessly, they often encourage passive acceptance over active knowledge transformation, reinforcing ``knowledge telling'' behavior --- a strategy where writing is a direct retrieval of information rather than formulated through deep reasoning~\cite{scardamalia1987knowledge}. The generative writing phenomenon raises urgent concerns about the homogenization of thought, erosion of writer ownership, and diminished self-regulation and reflection in writing~\cite{geng2024impact,peterson2025ai}. If generative-AI-powered writing tools primarily function as text generators, students may bypass critical stages in meaning construction, disengaging from essential cognitive processes that define effective and intentional writing. 

\begin{figure}[t]
    \centering
    \includegraphics[width=1\textwidth]{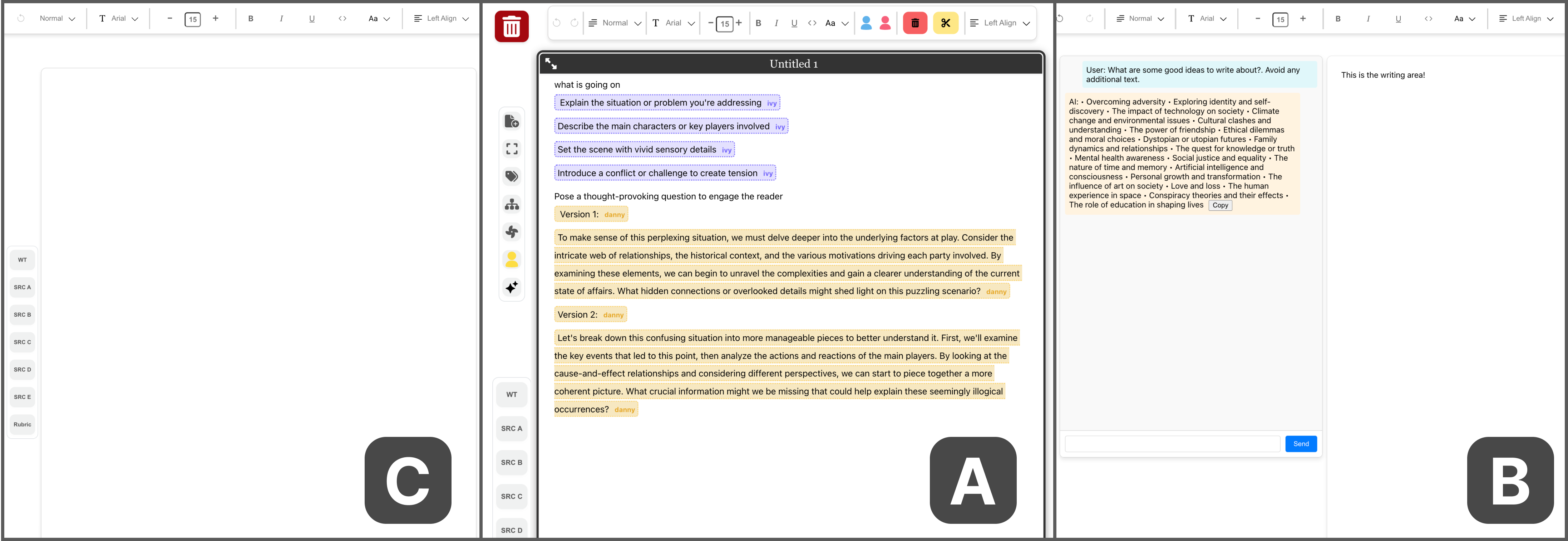}
    \caption{These are the three interface conditions for the experiment. [A] is \system; [B] is the chat-based interface condition; [C] is the standard interface. Each condition has an always-present dock that contains information about the source-based writing task. }
    \label{fig:codes}
\end{figure}

To address the problem of cognitive disengagement in AI-assisted writing, we need writing tools that take a process-oriented approach, guiding writers through key cognitive steps in writing rather than bypassing them. This approach augments human writers by structuring AI interactions to scaffold key writing processes --- brainstorming, argument development, content organization, rhetoric, revision, and reflection --- ensuring that AI enhances rather than replaces cognitive effort. Building on this perspective, in prior work, we developed a process-oriented writing tool called \system~\cite{siddiqui2025scriptshift} that supports writers through distinct phases of content development, revision, and rhetorical organization. Specifically, the tool allows 
writers to modularly structure their writing using a layered interface, invoke specialized AI assistants (Writer’s Friends) for targeted support, and seamlessly reorganize content to explore different rhetorical strategies. These affordances ensure that AI assistance is embedded within the writing process rather than disrupting it. In contrast, chat-based LLM interfaces provide AI assistance through linear, turn-based exchanges, requiring writers to manually integrate suggestions into their text. For example, when a writer receives paragraph-level feedback in a chat interface, they must switch contexts between the chat window and their document, manually copy-paste content, and then adjust the surrounding text for coherence --- a series of interruptions that fragments the cognitive flow of writing. This often disrupts the writing flow and encourages passive text adoption.

To systematically examine \textbf{how these differing AI interaction paradigms impact writer agency and knowledge transformation}, we conducted a randomized controlled trial (RCT) comparing \systems process-oriented design with a conventional chat-based LLM interface. This study evaluates how interface design shapes cognitive engagement, the depth of knowledge transformation, and writers' ability to maintain control over their composition process. Our study involved 90 undergraduate students randomly assigned across conditions (30 per condition). Participants complete a source-based writing task requiring synthesizing the provided documents into an essay arguing their chosen stance. In line with the focus of this paper, we focused on two constructs: 1) \textbf{knowledge-transformation:} Operationalized qualitative coding for markers of knowledge-transformation in the final essays, and 2) \textbf{writer agency:} Assessed through self-reporting for writing self-efficacy and perceived ownership.

Our findings reveal that writers experience significantly greater agency and demonstrate more markers of knowledge transformation when using process-oriented tools compared to chat-based interfaces. This aligns with theoretical predictions: process-oriented tools maintain separate content and rhetorical spaces, preserving the cognitive moves that characterize expert writing while providing targeted assistance at each stage. In contrast, chat-based interfaces often collapse these spaces by generating complete text, encouraging passive acceptance rather than active transformation. These results reinforce emerging evidence that conversational LLMs, while powerful, may not optimally support educational writing objectives when deployed through chat interfaces. Instead, we advocate for process-oriented AI design—systems that function as ``critical partners'' rather than text generators. By embedding LLMs within structured interfaces that scaffold (but do not automate) writing sub-processes, educators can empower students to engage deeply with content, refine their ideas, and maintain ownership of their work—key principles for effective human-AI collaboration in education. Enabling through granular scaffolding for diverse writing processes, we observe greater agency and deeper knowledge transformation for the process-oriented writing system than the alternative conditions.
\section{Related Works}

\subsection{Writing and Knowledge Transformation} 
Writing is a dynamic, non-linear process that cycles between planning, translating, and reviewing phases, with writers continuously adjusting their goals based on insights gained during composition~\cite{hayesflowercognition}. 
Scardamalia and Bereiter~\cite{scardamalia1987knowledge} developed the knowledge-transforming model of writing, in which content generation emerges from a dialectical process \cite{mcnamara2017toward} between two problem spaces: content (addressing belief and logical consistency) and rhetorical (handling compositional goals).
Novice writers, who lack the metacognitive skills to traverse this complex problem-solving model, default instead to the task-execution model of knowledge-telling where they retrieve information from memory based on genre and topic cues. Exercises like source-based writing tasks, emphasizing source understanding and idea synthesis, are powerful tools for evaluating and developing knowledge-transformation tendencies~\cite{segev2004writing}, which correlates with high-quality writing~\cite{rakovic2021automatic}. However, the rise of LLMs introduces complexities regarding writing development. While some evidence suggests AI assistance can yield immediate productivity gain~\cite{noy2023experimental}, other research highlights potential drawbacks. For example, reliance on current chat-based LLMs might undermine knowledge-transformation as writers often passively accept their suggestions to produce subpar and generic writing rather than engaging with them~\cite{tarchi2024use}.

\subsection{Writing and Agency} 
In the context of writing, agency refers to students' active, independent participation~\cite{dahlstrom2019digital} Writers achieve agency through engagement with specific contexts like classroom narrative writing~\cite{biesta2006agency,emirbayer1998agency}. LLMs raise concern among creative writers regarding ownership~\cite{gero2022sparks} and agency over content~\cite{hoquehallmark,mieczkowski_hancock_2022,mirowski2023co}, along with broader ethical and societal risks~\cite{weidinger2021ethical}. When using ChatGPT for source-based writing tasks, Tarchi et al. found reduced writer agency and a disconnect between source documents and the composed prose~\cite{tarchi2024use}. Chat-based LLMs homogenize content, hindering students' stylistic development~\cite{anderson2024homogenization}. While language is inherently dialogic—each utterance responding within broader sociocultural contexts~\cite{bakhtin2010dialogic}—current human-AI writing interactions fail to position voices effectively within these interconnected dialogues.

\subsection{AI and LLM in Writing Education} 
The integration of AI in writing education has roots in cognitive science research on computer text-editors as writing aids. Collins and Gentner~\cite{collins1980framework} emphasized how writers manage multiple cognitive processes throughout the writing process, and that computers are suited to supporting the cognitive processes they outline. Building on this foundation, early systems like iSTART~\cite{mcnamara2004istart} demonstrated computer-based scaffolding of metacognitive strategies through interactive training and pedagogical agents. Key findings revealed that computer experiences combining reading comprehension and writing strategy training enhanced students' knowledge-transformation abilities in source-based writing~\cite{weston2018comprehension}.

Recent advances have expanded both theory and application, particularly for knowledge-transformation and cognitive load management. There has been significant progress with machine learning algorithms for automatically identifying knowledge-transforming in argumentative essays~\cite{rakovic2021automatic}, while AI has demonstrated effectiveness in improving academic argumentation writing through AI-Supported Scaffolding (AISS) systems \cite{kim2022learner}. Studies reveal nuanced patterns of human-AI interaction: open-source LLMs perform comparably to proprietary models in providing writing feedback \cite{gubelmann2024exploring}, while research on real-time assistance tools shows writers engaging thoughtfully with AI suggestions rather than accepting them passively \cite{cummings2024unexpected}.  More saliently, Parker \cite{parker2024negotiating} showed that AI integration fosters metacognitive reflection and maintains writer agency through dynamic evaluation and negotiation.

However, several research gaps persist: (1) developing more sophisticated mechanisms for multilingual writing support and macro-level feedback that goes beyond surface corrections \cite{strobl2019digital}; (2) creating structured frameworks to accommodate diverse learning approaches while maintaining academic integrity, particularly as institutions adapt to ethically integrate AI technologies \cite{parker2024negotiating}; (3) improving AI algorithms specifically tuned for educational purposes rather than general text processing \cite{kim2022learner}; and (4) establishing robust validation methods for measuring educational impact across different writing contexts and complexities. 
  
\section{Methodology}
Our study's aim was to validate two constructs: Knowledge Transformation and Agency, within the context of different LLM-based writing scaffolding experiences. We structured the experiment as a randomized control trial with three conditions for writing: (1) \system, which provided integrated process-oriented LLM scaffolding; (2) a custom chat-based LLM to support writing; and  (3) a standard writing interface with no LLM support as a control to ground our findings. For the LLM conditions we used Claude 3.5 sonnet by Anthropic, known for its writing ability~\cite{claude}. We conducted this human subjects research under the purview of Stanford University's IRB. We recruited participants through Prolific~\cite{prolific}, with filtering criteria requiring them to be students actively pursuing undergraduate degrees in the United States. The participants consented to volunteer 1.5 hours for \$15. In addition to the main task, we asked participants to complete a pre-test survey for demographic and main task-specific questions, and a post-test survey geared towards understanding their writing experience. We did not collect personally identifiable information from participants.

\subsection{Participants}
We recruited 90 participants (Female= 47, Male= 42, Non-Binary=1) and randomly assigned them to one of three conditions (participant per condition=30). Participants reported age between 18 and 41, the mode of age range was 18-39 and the median age range was 26-30. The English proficiency of participants was high as the recruitment pool was restricted to university students in the United States (native=76, proficient=13, intermediate=1). All participants besides 3 had previously used AI for writing, most participants had used multiple tools (ChatGPT: 80, Grammarly: 55, Gemini:29, Claude:13, Others:11). The self-reported confidence of participants is as follows: \system ($\mu = 3.9$, $\sigma = 0.75$), chat-based interface condition ($\mu = 4.03$, $\sigma = 1$) and standard condition ($\mu = 4.06$, $\sigma = 0.72$). The median time for task completion was 1 hour 13 minutes. 

For the main writing task, we gave participants a source-based writing task to construct a position on \textit{advertisement} based on provided documents in 800-1000 words. We derived our writing task from a College Board AP Language and Composition exam due to its psychometric robustness. Before starting the main writing task, we showed participants an interface tutorial and assessed their understanding through a quiz. We provided participants with a grading rubric to guide their essay writing and additionally recommended composing their writings into three sections: introduction, body, and conclusion. We prescribed having three body paragraphs but made it clear that we accepted shorter essays if they were able to make their case sufficiently. We informed participants that using aid outside the interface was not allowed and that we tracked their interface transactions. We validated their submissions by assessing their interface logs.

Two independent raters assessed each essay using two evaluation frameworks: (1) using the AP Language and Composition essay rubric; (2) using a knowledge transformation evaluation, following the coding methodology from prior work~\cite{rakovic2019towards}. To motivate high-quality writing in our participants, we offered a \$5 bonus if their essay scored in the top 5 for their condition based on the AP rubric scores and the agreement between both raters. The protocol involved coding for knowledge-transforming and essay quality scoring including the raters independently coding a single essay and discussing their disagreements. Following this, each coder separately coded the rest of the essays. At the end of this first round, the inter-rater reliability was calculated using pooled Cohen's Kappa yielding a score of $\kappa = 0.76$. Following a discussion to resolve disagreements, raters re-coded for a score of $\kappa = 0.86$, which is a very good inter-rater reliability. Any remaining disagreement was resolved by an independent third rater.  

Following the writing task, a post-test survey was administered to assess participants' experience with their task. For participants in treatment conditions, we asked about their experience using the LLM interface and their perceived agency over their writing, allowing us to validate our agency construct.

\begin{figure}[t]
    \centering
    \includegraphics[width=1\textwidth]{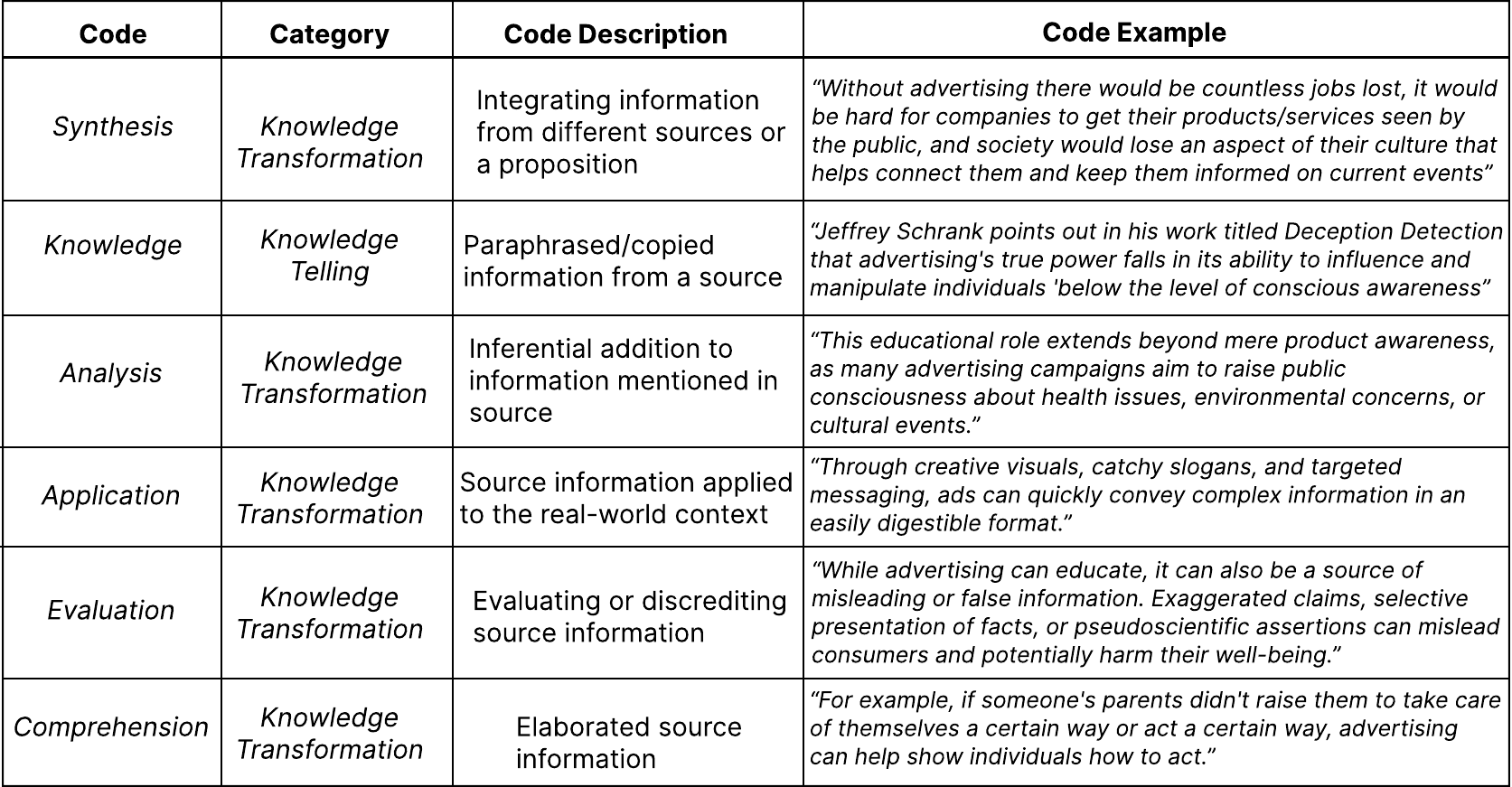}
    \caption{Codes For Knowledge Transformation (Synthesis, Analysis, Application, Evaluation, and Comprehension) and Knowledge Telling (Knowledge) from Raković et al. 2019 \cite{rakovic2019towards} with an example from our participant essays' coding.}
    \label{fig:codes}
\end{figure}

\section{Results}

\subsection{Knowledge transformation} To examine differences in knowledge transformation across conditions, we conducted Kruskal-Wallis tests followed by pairwise Mann-Whitney U tests. The Kruskal-Wallis test revealed significant differences in Analysis scores across conditions ($H = 8.72$, $p = .013$). Subsequent pairwise comparisons showed that \system participants displayed significantly higher Analysis performance compared to both Standard ($U = 236.50$, $p = .007$) and chat-based conditions ($U = 271.50$, $p = .020$). Additionally, \system participants showed significantly better performance in \textit{Evaluation} ($U = 287.50$, $p = .033$) and \textit{Knowledge} ($U = 554.00$, $p = .038$) compared to the ChatLLM condition. While no significant differences were found in \textit{Application} ($H = 0.67$, $p = .715$), \textit{Comprehension} ($H = 0.92$, $p = .632$), or \textit{Synthesis} ($H = 2.75$, $p = .253$), the overall pattern suggests that \system was more effective at facilitating higher-order knowledge transformation, particularly in analytical and evaluative tasks. These findings indicate that \system's structured approach may better support students in developing advanced cognitive skills compared to both traditional methods and standard AI writing assistance. In Figure~\ref{fig:survey_plots}, we showcase participant self-reporting for various knowledge transformation processes~\cite{scardamalia1987knowledge,scardamalia1984teachability}.

\subsection{Agency} The post-test survey contained 12 questions on a 7-point Likert scale, ranging from Strongly Disagree to Strongly Agree, for the chat-based and \system condition. Due to the nature of the control condition, there were no questions regarding the perception of AI usage as the system and only three questions regarding writing agency for establishing a baseline to compare the two AI conditions. The questions were: (Q1) \textit{``I felt in control during the writing process''}, (Q2) \textit{``I feel content with the writing I produced''} and (Q3) \textit{``I would feel comfortable publishing this essay in my name.''} One-way ANOVAs revealed significant differences between conditions for all three measures: \textit{Q1} ($F(2, 85) = 101.90$, $p < .001$), \textit{Q2} ($F(2, 85) = 72.91$, $p < .001$), and \textit{Q3} ($F(2, 84) = 26.15$, $p < .001$). 

To examine specific differences between the \system condition and other conditions, we used Welch's t-test due to its robustness against unequal variances. For \textit{Q1}, \system ($\mu = 6.59$, $\sigma = 0.87$) showed significantly higher ratings compared to both chat-based ($\mu = 3.34$, $\sigma = 0.72$; $t(54.20) = 15.48$, $p < .001$) and Standard conditions ($\mu = 5.83$, $\sigma = 1.09$; $t(55.06) = 2.95$, $p < .01$). Similar patterns emerged for \textit{Q2}, where \system ($\mu = 6.34$, $\sigma = 1.11$) demonstrated significantly higher satisfaction than chat-based ($\mu = 2.83$, $\sigma = 1.04$; $t(55.74) = 12.46$, $p < .001$) and Standard conditions ($\mu = 5.50$, $\sigma = 1.31$; $t(56.10) = 2.68$, $p < .01$). For \textit{Q3}, \system ($\mu = 5.90$, $\sigma = 1.54$) also showed significantly higher comfort levels compared to chat-based ($\mu = 3.21$, $\sigma = 0.82$; $t(42.60) = 8.29$, $p < .001$) and Standard conditions ($\mu = 4.90$, $\sigma = 1.76$; $t(55.06) = 2.30$, $p < .05$). 

\begin{figure}[t]
    \centering
    \includegraphics[width=1\textwidth]{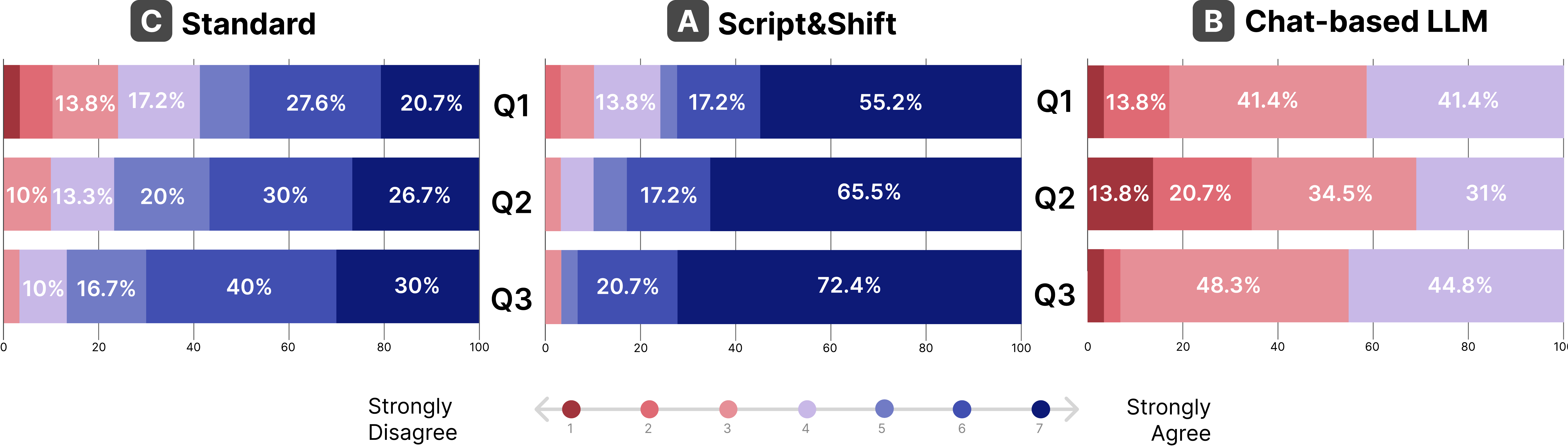}
    \caption{This figure compares agreement for (A) \system, (B) Chat-based LLM, and (C) Standard interfaces across three dimensions of agency: perceived control during writing (Q1), content satisfaction (Q2), and publication comfort (Q3).}
    \label{fig:agency}
\end{figure}

\begin{figure}[t]
    \centering
    \includegraphics[width=1\textwidth]{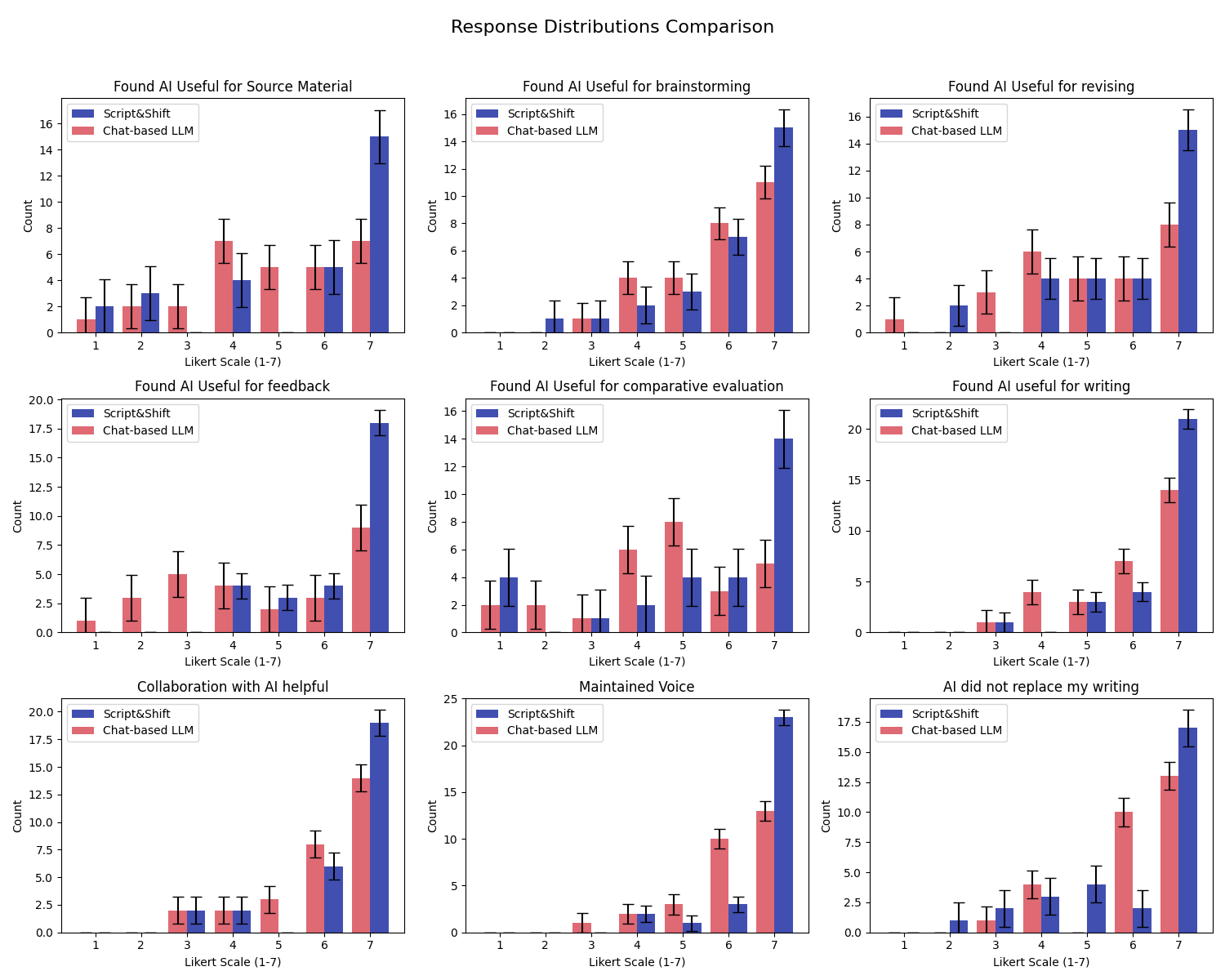}
    \caption{These plots are for individual questions in the survey on a 7-point Likert scale to measure writing scaffolding and agency experienced by the writer.}
    \label{fig:survey_plots}
\end{figure}

Across all measures, the differences between \system and chat-based conditions were notably larger than those between \system and Standard conditions ($\Delta\mu_{Shift-Chat} > \Delta\mu_{Shift-Standard}$), suggesting that \system substantially improved the writing experience compared to both baseline and chat-based approaches, look at Figure~\ref{fig:agency} for more details. To isolate the perception of AI with respect to agency, the following two items were present in the case of \system and chat-based interface: (1)\textit{ ``I felt that my essay maintained my original voice even when using AI''} and (2)\textit{ ``I felt that the AI did not replace my writing.''} For the first question (see Figure~\ref{fig:survey_plots} for more details), \system ($\mu = 6.62$, $\sigma = 0.86$) showed marginally higher ratings compared to the Chat condition ($\mu = 6.10$, $\sigma = 1.08$; $t(53.38) = 2.01$, $p < .05$). However, for question 2, there was no significant difference between \system ($\mu = 5.97$, $\sigma = 1.52$) and Chat conditions ($\mu = 6.07$, $\sigma = 1.18$; $t(55.88) = -0.29$, $p > .05$). These results suggest that while \system maintained slightly higher perceived effectiveness for question 2, both conditions performed similarly on this measure.

\subsection{Essay Scoring} While the \system condition showed the highest College Board scores ($\mu = 5.34$, $\sigma = 2.75$), the substantial variability across Standard ($\mu = 5.04$, $\sigma = 1.82$) and ChatLLM ($\mu = 4.52$, $\sigma = 2.59$) conditions suggests individual differences may have outweighed experimental effects. Source Integration revealed marginal differences between \system ($\mu = 3.31$, $\sigma = 1.78$), Standard ($\mu = 2.75$, $\sigma = 1.27$), and ChatLLM ($\mu = 2.83$, $\sigma = 1.53$) conditions, though high variance indicates considerable group overlap. Other dimensions were similar across the board with moderate variability.

\subsection{Correlation Between Interface Use and Knowledge Transformation} 
\begin{figure}[t]
    \centering
    \includegraphics[width=0.9\textwidth]{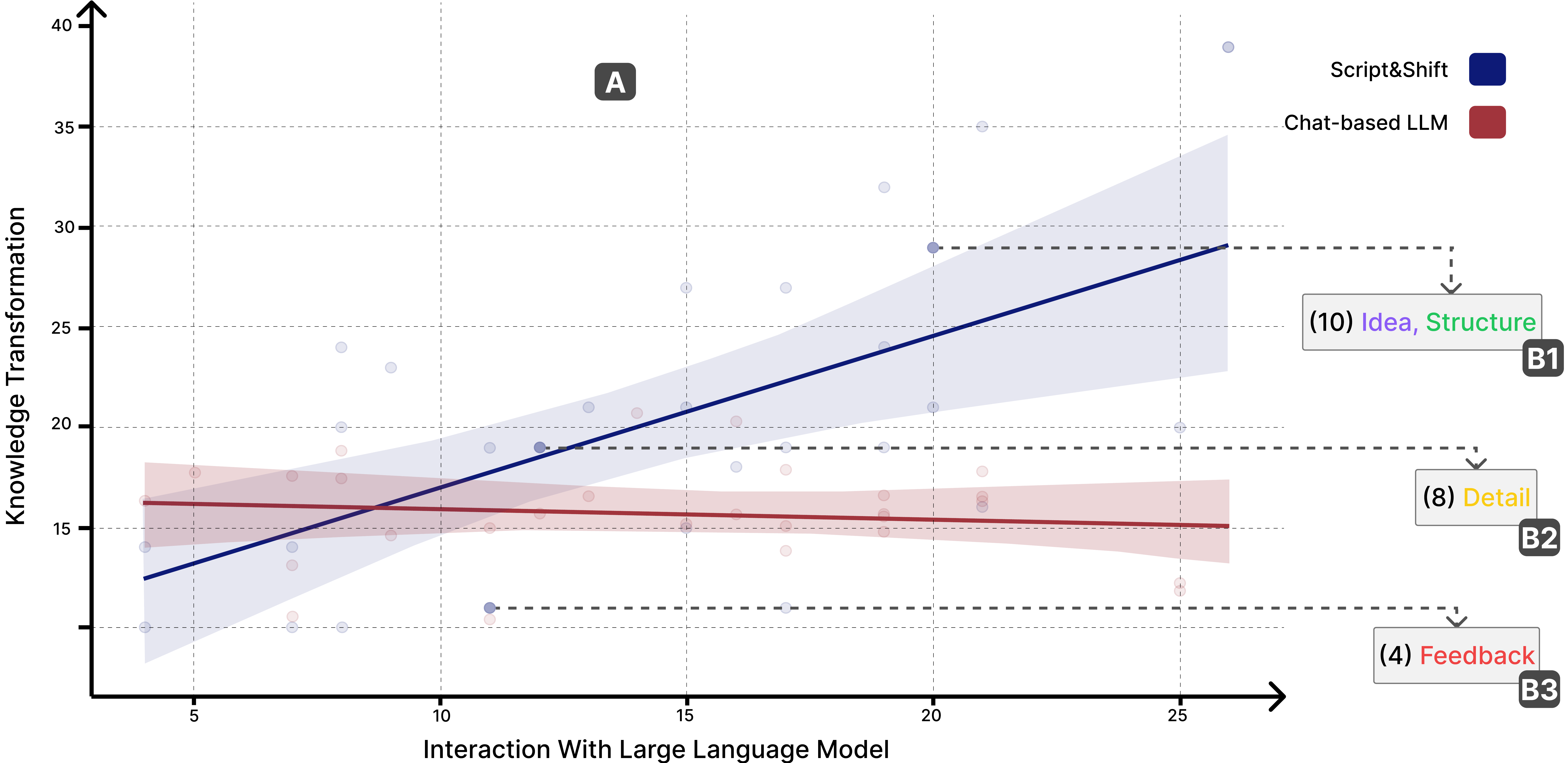}
    \caption{(A) shows the correlation between knowledge transformation markers and LLM feature access. While chat-based shows no correlation, \system exhibits moderate positive correlation ($r = 0.608$, $p = 0.001$). (B1-B3) display most-used LLM features for three representative \system participants.}
    \label{fig:transform}
\end{figure}

Analysis revealed a moderate positive correlation ($r = 0.61$) between interface transactions ($\mu = 14.43$, $\sigma = 5.95$) and knowledge transformation markers ($\mu = 15.29$, $\sigma = 7.40$). The relatively high standard deviations in both measures suggest substantial variability in how participants engaged with the interface and transformed knowledge. Figure~\ref{fig:transform}A shows the correlation between interface use and knowledge transformation. Among participants in the \system condition with high knowledge transformation, \textit{`Idea Ivy'} (Fig.~\ref{fig:transform}B1), the AI supporting brainstorming, was used most frequently. We observed that participants with moderate knowledge transformation favored \textit{`Detail Danny'}(Fig.~\ref{fig:transform}B2), while those with low knowledge-transformation and overall lower LLM usage primarily relied on feedback features, \textit{`Audience Ali'} and \textit{`Feedback Felix'} (Fig.~\ref{fig:transform}B3), that do not allow writers to explicitly specify their prompts. This usage pattern mirrors our prior findings for interfaces without \systems spatial affordances~\cite{siddiqui2025scriptshift}, suggesting participants with low knowledge-transformation did not utilize the spatial features of the interface available to them.

\subsection{Prompt Characteristics.} 
We analyzed participants' prompts across both LLM interfaces by categorizing their \textit{target} (T), \textit{constraint} (C), and \textit{output} (O). Evaluation-focused prompts  $(Text_T, Quality_C, Evaluate_O)$ were preferred by high knowledge-transformation participants. In \system, they also utilized generation and organization prompts, $(Text_T, Generate_C, Evaluate_O)$ and $(Structure_T, Organize_C, Revise_O)$. In the chat-based interface condition, they preferred generation with examples $(Text_T, Generate_C, Example_O)$. On the other hand, low knowledge transformation participants mainly used elaboration $(Text_T, Elaborate_C, Expand_O)$ in \system and generation prompts $(Text_T, Generate_C, Expand_O)$ in chat-based interfaces. 

\section{Discussion}
Our study findings indicate that using an integrated process-oriented AI writing tool like \system enhances writers' agency over the writing process compared to chat-based writing tools. Notably, the self-reported agency was also higher for users in \system condition than for writers in the control condition, where they worked with complete autonomy (no AI assistance). An explanation for \system fostering greater agency than even complete autonomy could be, that without any AI mediation, subjects in the control condition likely had a different mental model of what ``control'' is, e.g., how they felt about the writing prompt. 

Participants were more comfortable publishing work from \system ($\mu = 5.90$) versus the chat-based ($\mu = 3.21$) and standard interface conditions ($\mu = 4.90$). This suggests process-oriented scaffolding better preserves writers' sense of ownership than tools that write for them, supported by \system's higher ratings for maintaining original voice. Such interfaces provide diverse combinations of \textit{target}, \textit{constraints}, and \textit{outputs} in prompts, facilitating reflective processes that support knowledge-transformation~\cite{scardamalia1987knowledge,scardamalia1984teachability}.

The correlation results between LLM usage and knowledge-transformation provide some support for \system's process-oriented design approach. The moderate positive correlation ($r = 0.61$) suggests some relationship between active engagement with the tool and depth of cognitive processing. The substantial variation in both interface transactions ($\mu = 14.43$, $\sigma = 5.95$) and knowledge-transformation ($\mu = 15.29$, $\sigma = 7.40$) markers complicates this interpretation, but it may also indicate that our participants engaged in different strategies while using the tool and in knowledge-transformation during writing. \system demonstrated higher overall and component-specific knowledge-transformation trends, with its integrated support fostering deeper content engagement versus the passive acceptance of shallow insights observed in the chat-based condition. 

While we were unable to discriminate between essay quality across the three conditions, despite greater markers of knowledge-transformation in \system condition, it is not unexpected. Prior studies that have analyzed essays for markers of knowledge-transformation and graded them have found a weak correlation at best~\cite{rakovic2021automatic}. This disconnect may reflect the complexity of writing quality assessment. Grading by The College Board rubric shows the highest means for \system ($\mu = 5.34$), but with substantial variability ($\sigma = 2.75$), suggesting that individual differences may have played a larger role than AI tool support in determining final essay quality. To observe a significant effect size, longer-term treatments have to be applied. Several limitations should be considered. The 1.5-hour timeframe may have constrained writers' ability to fully engage in knowledge-transformation and deeper cognitive processes.

\section{Limitations \& Future Work}
Our study revealed knowledge-transformation trends in single writing samples produced after experimental exposure to new technologies. \system demonstrated promising patterns in knowledge-transformation and writer agency, suggesting directions for future AI-assisted writing research.

We aim to explore how LLM-based systems can augment pedagogical strategies to produce better learning outcomes for students. While prior research indicates unintended negative consequences of ChatGPT on student writing~\cite{tarchi2024use}, we believe integrated process-oriented systems can amplify students' abilities while aligning intent with process-specific scaffolding~\cite{parker2024negotiating}.

Despite their potential, the risks of hallucination and bias remain serious. While these systems can help students become better writers, they may also produce a homogenizing effect~\cite{anderson2024homogenization}. This can lead to the propagation of biases and misinformation encoded in these systems. Additionally, school personnel are concerned about the impact of generative technologies in classrooms~\cite{bowman2022chatbot,meyer2023chatgpt}, as students often use them to complete assignments while forgoing crucial learning steps. Under such circumstances, it becomes imperative to build technologies that can be monitored by teachers to provide students with safe access to these technologies, which could otherwise have severe unintended consequences.
\section{Conclusion}
This paper shares findings from a randomized control trial comparing chat-based writing interface with integrated process-oriented writing tools like \system for the constructs of knowledge-transformation and experienced agency. The support provided by \system for varying writing processes helped writers map their intentions to different scaffolds to produce writing rich in rhetorical and content interplay. We believe technologies like these can help improve student writing while empowering them to engage with their content deeply.

\section{Acknowledgment}
We are grateful to the reviewers for their helpful feedback. We also thank Jessica Roberts for her help with the statistical analysis.  This work is supported through the AI Research Institutes program by the National Science Foundation and the Institute of Education Sciences, U.S. Department of Education through Award $\#2229873$ - National AI Institute for Exceptional Education.

\bibliographystyle{splncs04}
\bibliography{99_refs}
\end{document}